\documentclass[prl,floatfix,twocolumn]{revtex4}
\usepackage{epsf}

\newcommand{\be}{\begin{equation}}
\newcommand{\ee}{\end{equation}}

\begin{document}

\title{Large $N$ reduction in the continuum three dimensional Yang-Mills theory}
\author{R. Narayanan}
\affiliation{
Department of Physics, Florida International University, Miami,
FL 33199}
\author{ H. Neuberger}
\affiliation{
School of Natural Sciences,
Institute for Advanced Study, Princeton, NJ 08540\\
Rutgers University, Department of Physics
and Astronomy,
Piscataway, NJ 08855 \footnote{Permanent Address.}
}

\begin{abstract}
Numerical and theoretical evidence leads us to propose the following: 
Three dimensional Euclidean Yang-Mills theory in the planar limit
undergoes a phase transition on a torus of side $l=l_c$.
For $l>l_c$ the planar limit is $l$-independent, as expected of  
a non-interacting string theory. We expect the situation 
in four dimensions to be similar. 
\end{abstract}

\maketitle

{\bf Introduction.}
Yang-Mills theory in three dimensions is similar to Yang-Mills theory
in four dimensions in exhibiting a positive mass gap, linear confinement,
a finite temperature deconfinement transition and a sensible large $N$ 
limit~\cite{3d}.

Nonabelian gauge theories in 3 and 4 dimensions interact strongly at large
distances and weakly at short distances. Doing calculations that bridge
these two regimes remains a major challenge, of central importance to particle
physics. It has been a long held hope that the task would simplify at infinite
number of colors, $N$. Here, at strong coupling,  a fitting hypothesis is that
the theory describes free strings, while at 
weak coupling, the theory certainly describes
weakly interacting particles. The result of this paper indicates that strong and 
weak coupling are
separated by a phase transition at infinite $N$. 
Specifically, we provide numerical evidence
that $SU(N)$ gauge theory on a three torus of side $l$ undergoes a transition
at a critical length, $l=l_c$. For any finite value of $N$ there cannot be any 
phase transitions in this system. The existence of such a transition at 
infinite $N$ is surprising, raises questions about the usually assumed smooth
dependence of observables on momenta and might indicate deeper connections
between gauge theory, string theory and random matrix theories. 

Over twenty years ago, in the context of $SU(N)$ lattice gauge theory,  
Eguchi and Kawai~\cite{ek} made the observation that at infinite 
number of colors space-time can be replaced by a single point. 
This dramatic reduction in the number of degrees of freedom 
should make it easier to deal numerically with planar QCD
than with ordinary, three color QCD. For a practical procedure 
it is essential that some version of large
$N$ reduction also hold in the continuum, not just on the lattice.  
Previous attempts to define a continuous reduced
model had problems with topology and fermions~\cite{gk}. 

We focus on the continuum limit 
of pure lattice YM defined on a torus
and try to establish that 
expectation values of traces of Wilson loop 
operators do not depend on the size of the torus. Wilson loops
of arbitrary size can be folded up into the torus and correctly
reproduced~\cite{t3}.
The
lattice is essential because it provides a regularization with
well defined loop equations~\cite{mm}. Loop equations provide a
convenient tool to establish reduction. 

We restrict ourselves to three dimensional theories for
numerical reasons.
We find that continuum large $N$ 
reduction holds so long as
the torus is large enough.
The critical side length
of a symmetrical torus is denoted by $l_c$ and is defined 
in terms of a microscopic fundamental physical scale
of the theory. Solving the theory for some $l>l_c$ would produce
complete and exact 
information at leading order in $N$ for any $l$. 
The system as a whole undergoes a phase transition at $l=l_c$.
The number of sites in a numerical simulation 
in a given
direction, $L$, determines the maximal value the ultraviolet cutoff $\Lambda$ 
can take. It is $\Lambda=\frac{L}{l_c}$. 
For a Wilson action the lowest
$L$ that has some semblance to continuum is $L=3$. Thus, at the expense of 
larger $N$ one can get numerically close to continuum using very 
small lattices. The values of $N$ needed are of order $20$ to $50$ 
and this trade-off is worth taking. 

If a similar result holds in four dimensions, a shortcut
to the planar limit becomes a realistic option. Our experience makes
us hopeful and our tools should allow us to tackle four dimensions in 
the future.

{\bf A lattice argument.}
There is a global 
$Z^d(N)$ ($U^d(1)$ in the $N\rightarrow\infty$ limit) 
symmetry on the torus that leaves
contractible Wilson
loop operators invariant but multiplies
Polyakov loops winding around a direction $\mu$ 
by a phase $e^{\frac{2\pi\imath}{N}k_\mu }$.
The preservation of this symmetry
is crucial for large $N$ reduction~\cite{bhn}. 
Eguchi and Kawai have shown that the lattice loop equations in the 
$N=\infty$ limit on a single site lattice are the same as on 
an infinite four dimensional lattice as long as the
$U^4(1)$ symmetry is unbroken. 
The continuum limit in the single site lattice model
has to be taken by sending the lattice coupling
$b=\frac{1}{g^2 N}$ to infinity, but in $d>2$ 
a phase transition occurs, blocking the way. 
At the transition the $Z^d(N)$
symmetry breaks spontaneously, ruining the equivalence of
loop equations. 
It is possible to fix the single site lattice model by
quenching~\cite{gk,bhn} 
or twisting~\cite{twist} the system.
We take a different approach here.
The proof of Eguchi and Kawai
goes through for a lattice torus of size $l_1\times l_2...$
with arbitrary $l_\mu$ and in any dimension. The loop equations,
together with boundary conditions for small loops, establish 
equality of expectation values of traces of operators associated
with arbitrary finite closed loops in the infinite volume theory
and their folded, contractible, counterparts on the torus.
Suppose we 
reduced the model to only an $L^d$ lattice with $L>1$: 
Again we expect the global
symmetry to break if $b>b_c (L)$ and reduction will hold for
$b<b_c(L)$. $b_c (L)$ 
will increase with $L$ and if 
$b_c(L)$ depends asymptotically on $L$ as dictated by
microscopic scaling for $d=3,4$ then continuum large $N$ reduction
will hold if we take the limit by keeping $b<b_c(L)$ and taking
$b\rightarrow\infty$.

In the approach pursued here,
we have to deal with one lattice artifact.
There will be a cross-over in the lattice internal energy
for the Wilson gauge action
at some small $b$ for a finite torus size and a finite
$N$. 
The cross-over becomes 
a ``bulk'' transition at infinite $N$, occurring at $b^B_c(N=\infty, L)$
for any finite lattice of size $L^d$ 
in lattice units.
Lattice large $N$ reduction would imply that $b^B_c(N=\infty, L)$
does not depend on $L$, $b^B_c(N=\infty, L)={b^B_c}^\infty$.
This is consistent with numerical simulations. 
The loop equation, together with constraints which come from
the parallel transporters being unitary matrices, produce 
the ``bulk'' transition without loosing their validity or changing their
form. The lattice transition occurs when the unitary matrix associated with
the one by one loop
opens a gap at eigenvalue -1 in its spectrum in the large $N$ limit. 
As $b$ increases further the gap widens. 
In the continuum this means that parallel transport round a tiny
loop will not differ much (in norm) from the identity. 
Similar transitions occur at $b^B_c(N,L=\infty)$ 
for large enough $N$. 
The common limiting value 
at $b^B_c(N=\infty, L=\infty)={b^B_c}^\infty$ is  
rapidly approached. This family of transitions are lattice
artifacts not associated with any symmetry breaking.
Examples are the Gross-Witten~\cite{gw} transition in two dimensions and 
Creutz's transitions~\cite{creutz} for $N>4$ in four dimensions. 

Even though lattice reduction is valid on either side of ${b^B_c}(N=\infty, L)$
as long as one is below $b_c(L)$, we have to be above
${b^B_c}(N=\infty, L)$ to realize continuum reduction.
For $L=1$ (the Eguchi-Kawai model) and
$d > 2$, the infinite $N$ ``bulk'' and 
$Z^d (N)$ breaking transitions accidentally fuse at a $b_c\ne{b^B_c}^\infty$. 
Similar ``accidents'' can happen for L=2,3.., but 
a window opens for large enough $L$ between ${b^B_c}^\infty$ 
and $b_c(L)$.
In three dimensions,
an $L=3$ lattice already has a window.

\begin{figure}
\epsfxsize = 0.45\textwidth
\centerline{\epsfbox{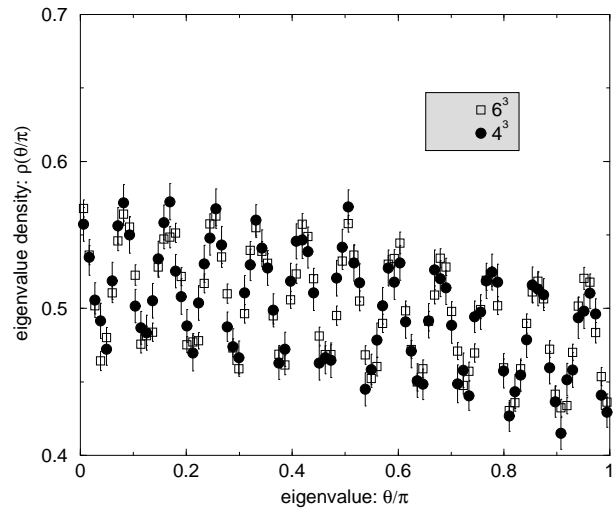}}
\caption{ Eigenvalue density distribution of a $4\times 4$ Wilson loop
on $4^3$(folded) and $6^3$(unfolded) at $b=0.66$ and $N=23$.
}
\label{fold}
\end{figure}

\begin{figure}
\epsfxsize = 0.45\textwidth
\centerline{\epsfbox{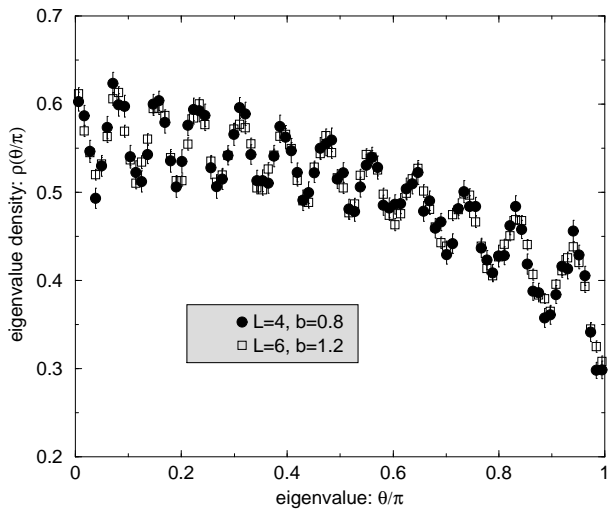}}
\caption{
Eigenvalue density distribution of $L\times L$ Wilson loop on
$L^3$ for $L/b=5$ and $L=4,6$. $N$ is set to $23$.
}
\label{scale}
\end{figure}

${b^B_c}^\infty=0.5$~\cite{gw} and $b_c(L)=\infty$ in $d=2$. 
The $U^2(1)$ symmetries are not broken and continuum
reduction works on tori of any size in two dimensions. 
The ``bulk'' transition occurs close to ${b^B_c}^\infty=0.4$
in $d=3$. 
Ordinary scaling in $d=3$ would require $\frac{L}{b_c(L)}$ to approach
a finite nonzero limit as $L\to\infty$.  Monte Carlo simulations were
performed using a combination of heat-bath updates by $SU(2)$ subgroups
and of full $SU(N)$ over-relaxation steps.
Ultraviolet fluctuations in loop observables were suppressed
by APE blocking~\cite{ape}.
We monitored the eigenvalue distribution of
the Polyakov loops in the three directions and found that
$0.6 < b_c(3) < 0.7$, $0.8 < b_c(4) < 0.9$, $1.0 < b_c(5) < 1.2$
and $1.2 < b_c(6) < 1.35$. When combined, these results indicate
that the scaled critical coupling $\frac{L}{b_c(L)}$ is in the
region $[4.2,5]$.
We compared folded and unfolded versions of the same loop
on tori of different sizes and found the spectral
densities associated with them to match as long as 
$\frac{L}{b(L)}\ge 5$.
An example of such a comparision is shown in Fig.~\ref{fold}.
We also checked scaling by comparing Wilson loops of same
physical size at different lattice spacings. An example of scaling
is shown in Fig.~\ref{scale}.

{\bf Continuum perturbation theory.}
If we had a scalar field theory where the field is a hermitian $N\times N$
matrix we know that independence on the torus size is impossible. This
dependence does not go away in the planar limit. On the level of Feynman
diagrams (taken in coordinate space) 
it is easy to calculate the dependence on torus
size for large $l$, in particular if the theory is massive~\cite{fz}. The leading
correction is exponentially suppressed in $l$ and comes from one virtual
particle going round a non-contractible circle on the torus. There is a
stable particle like that and it is in the adjoint representation of
$SU(N)$. In the gauge case, if there is confinement, we could use only
singlets under $SU(N)$ and, at infinite $N$, these singlets make sub-leading
contributions to the free energy at leading, ${\cal O}(N^2)$, order. 
We conclude
that for a confining gauge theory a 
planar diagram with a ribbon (double-line) 
representation of propagators makes no
contribution if one tears one of the propagators out of 
the surface and winds
it round the torus. 

Another way to see how reduction works in perturbation theory is to
understand what happens to momentum space~\cite{bhn}. Having a torus means that
momenta are quantized in units of $\frac{2\pi}{l}$ and there is no
way around this for a massive scalar matrix field. In the gauge
case the Feynman expansion starts from a constant gauge field background.
The gauge invariant content of this moduli space consists of $d$ sets 
of angles $\theta_\mu^i$ which effectively fill the intervals between
the quantized momenta making momentum space continuous and $l$
independent. The filling has to be uniform and this is true
at infinite $N$ if the global $Z^d(N)$ symmetry is unbroken.
The background - in a translation invariant gauge - is given 
by $A_\mu=diag(\theta_\mu^1,\theta_\mu^2,....\theta_\mu^N)$
but only the set $\{e^{\imath l\theta_\mu^1},e^{\imath l\theta_\mu^2},...
e^{\imath l\theta_\mu^N}\}$ labels truly distinct vacua. This is why
$|\theta_\mu^i|<\frac{\pi}{l}$, exactly as needed to fill in
the gaps between the $\frac{2\pi k}{l}$'s. At infinite $N$ the vacuum 
is characterized by $d$ eigenvalue distributions on the complex unit circle.

{\bf Hints from string theory.}
In view of developments during the last 
few years~\cite{strings} it seems more likely
now than ever before 
that indeed large $N$ $SU(N)$ pure gauge theories 
are equivalent to some string theory at zero string coupling. 
This means that the logarithm of the partition function defined
on a finite torus and divided
by the volume of the torus, is, in the planar limit, given
by a sum of extended, spherical, two dimensional excitations
embedded in the same torus. But, there is no way for the spherical surface
to become non-contractible on the torus and thus it cannot detect
that target space is a torus~\cite{aw}. Hence, one can have no dependence
on $l$. It is well known that simple string models on toroidal
backgrounds cannot distinguish very large radii from very small ones; 
$l_c$, as a minimal radius, realizes a similar phenomenon in the 
unknown non-interacting string theory describing planar three dimensional
pure YM. 

It used to be revolutionary to think that statistical field
theories on finite volumes can have phase transitions. This is no longer
true.
To the
early
toy model examples~\cite{toy} we can add now cases of true,
full fledged field theories with real relativistic degrees
of freedom, also developing 
phases transitions in the planar limit~\cite{strings}.

{\bf Large $N$ phase transitions.}
Large $N$ transitions may emerge as quite ubiquitous in
continuum gauge theories. There are transitions, like the
one presented in this paper, that affect the system as a whole,
but there are also other transitions that affect only a class
of observables \cite{do, dk, wsph, gm}. 

The basic observables used in our study have been the distribution of
eigenvalues of Wilson and Polyakov loops.  For large $N$ these
observables are unconventional because they involve traces of all
powers of the basic unitary matrix, not only a few low powers. Thus,
issues of renormalization require more work~\cite{ren}. If
these issues can be resolved, we might be able to 
exploit the fact that eigenvalues of large 
matrices have many universal
properties~\cite{mehta}: The dynamics of the gauge theory could be
encoded in the transformations one needs to carry out in order to
bring these eigenvalue distributions to universal forms. 
While there are difficulties in continuum perturbation
theory,
the situation on the lattice is very clear: We numerically
look for features that scale
as the universal features of the field theory would have it.

The simplest strong-weak transition would be associated with
Wilson loops: Small loops
will have parallel transporters with a spectral gap and big loops
will have almost uniform distributions to account for confinement
in all finite irreducible representations. 
At finite $N$ there are no gaps in the spectra but, in the
range of the would be gaps, the eigenvalue 
density is exponentially suppressed as $N$ increases.

{\bf Beyond the transition.}
For $l$ just a bit smaller than $l_c$ exactly one of the $Z(N)$
factors in the $Z^3(N)$ breaks spontaneously. Thus, the forty eight
element cubic symmetry group of our equal sided torus breaks down
to an eight dimensional group acting in the plane perpendicular
to the direction in which the Polyakov loop spectra took on non-uniform
structure.  

In order to prepare ourselves for what to
look for when the torus is further squeezed 
we studied the $1^d$ EK model, now
interpreted as a simple effective model for the dynamics
of the vacuum manifold of the full system. 
Simulations we have carried out in three and four dimensions
showed that at infinite $N$ these models undergo  
a staircase of transitions, breaking
one additional $Z(N)$ factor at a time. The possible continuum meaning 
of the various intermediate phases will have to wait for more work. 

In super-symmetric YM gauge theories, compactified super-symmetrically
on tori, the perturbative mechanism driving the spontaneous
breaking of the $Z^d(N)$ symmetry can be eliminated. 
Beyond perturbation
theory we do not know the answer, and other global symmetries come into 
play. It is conceivable that in some cases $l_c=0$, indicative of a 
pure matrix model representation
of the planar limit of a continuum gauge theory. 
Although 
the physical size is zero, regularization issues might require
one to take $L\to\infty$ in a way correlated
with $b\to\infty$, and the zero size model may not admit a definition
as the large $N$ limit of an ordinary matrix integral.

{\bf Future lattice work.}
Building on earlier two dimensional work we know how to calculate
meson propagators in the planar limit. Meson momenta of values
below the ultraviolet cutoff can be introduced by multiplying the
original link matrices $U_\mu (x)$ by phase factors $e^{\imath p_\mu}$.
The $p_\mu$ allow to tune the momenta carried by the mesons to
desired values. 
One has no finite volume effects to worry about: to get to the
continuum limit one just increases
$b$, making sure that $l$ stays larger than $l_c$. Values of $N$
in the range of few tens seem to be adequate. The lattice Dirac
matrices are much smaller (and much denser) than in usual simulations.
We would be able to address the smoothness of
the two point meson correlation function, at infinite $N$, 
as a function of $q^2$, where $q$ is Euclidean momentum.
Could there be a non-analyticity at some $q^2$? After all,
if the crossover between {\it physical} strong and weak gauge forces
happens in a range of scales that shrinks to zero at infinite
$N$, phase transitions may occur in every observable, not only
special ones, like Wilson loops. In four dimensions this could,
finally, bring about a peaceful coexistence between 
large $N$ and instantons~\cite{inst}.

In parallel simulations of the pure gauge case,
large $N$ work will require 
a floating point effort per node that grows
as $N^3$ while communication demands will only grow as $N^2$. So, PC farms
with off the shelf communications would be well suited.  

{\bf Conclusions.}
Our final conjecture about three dimensions is stated in
the abstract. We call it a ``conjecture'' because our
numerical tests have been relatively modest and because the consequences
of the conjecture could be far reaching: many applications
of 't Hooft's large $N$ limit~\cite{thooft} assume analyticity in momenta
and this assumption is now challenged.  
Our evidence is a combination of numerical work
and more theoretical observations. On the numerical side we see the
$Z^3 (N)$ symmetry breaking point on the lattice change with lattice
size in a way consistent with continuum scaling. Theoretically, 
lattice large $N$ reduction based on large $N$ loop equations 
is a strong coupling argument while averaging over the moduli
space of constant abelian connections at weak coupling resolves
an apparent contradiction with conventional wisdom about finite
size effects. 

{\bf Acknowledgments.}
We would like to thank Joe Kiskis with whom we collaborated in the initial
stages of this project. H. N. would like to thank M. Douglas, V. Kazakov, 
I. Klebanov, J. Maldacena, N. Seiberg, E. Witten for useful comments. 
R. N. acknowledges a contract from Jefferson Lab under which this
work was done. The Thomas Jefferson National Accelerator Facility
(Jefferson Lab) is operated by the Southeastern Universities Research
Association (SURA) under DOE contract DE-AC05-84ER40150.
H. N. acknowledges partial support at the Institute for Advanced Study
in Princeton from a grant in aid by the Monell Foundation, as well as
partial support by the DOE under grant number 
DE-FG02-01ER41165 at Rutgers University. 
Scientific Computing facilities at Boston University were used
for part of the numerical computations.


\begin{thebibliography}{99}
\bibitem{3d} D. Karabali, C. Kim, V. P. Nair, Nucl. Phys. B434 (1998) 103;
M. Teper, Phys. Rev. D59 (1999) 014512; S. Dalley, B. van de Sande, 
Phys. Rev. D63 (2001) 076004.
\bibitem{ek} T. Eguchi, H. Kawai, Phys. Rev. Lett. 48 (1982) 1063.
\bibitem{gk} D. J. Gross,  Y. Kitazawa, Nucl. Phys. B206 (1982) 440.
\bibitem{t3}
Closed finite 
loops in $R^d$ and contractible
loops on $T^d$ are simply related: 
The equivalence class of loops identifiable under translations
is defined by the tangent vectors $t_\mu(s)=\frac{dx_\mu}{ds}$ 
to the curve ${\it C}=\{x_\mu (s)\}$. The functions $t_\mu(s)$ 
reconstruct the class on a torus of any given size.
\bibitem{mm} Yu. M. Makeenko, A. A. Migdal, Phys. Lett. 88B (1979) 135. 
\bibitem{bhn} G. Bhanot, U. M. Heller, H. Neuberger, Phys. Lett B113 (1982) 47;
H. Levine, H. Neuberger, Phys. Lett. B119 (1982) 183, J. Kiskis, R. Narayanan, 
H. Neuberger, Phys. Rev. D66 (2002) 025019.
\bibitem{twist} A. Gonzalez-Arroyo, M. Okawa, Phys. Rev. D27 (1983) 2397.
\bibitem{gw} D. J. Gross, E. Witten, Phys. Rev. D21 (1980) 446.
\bibitem{creutz} M. Creutz, Phys. Rev. Lett.46 (1981) 1441.
\bibitem{ape} T. DeGrand, Phys. Rev. D63 (2001) 034503;
M. Albanese et. al. [APE Collaboration], Phys. Lett. B192, (1987) 163;
M. Falcioni, M.L. Paciello, G. Parisi and B. Taglienti, Nucl. Phys. B251 
(1985) 624. 
\bibitem{fz} H. Neuberger,  Phys. Lett. B233 (1989) 183.  
\bibitem{strings} O. Aharony, S. S. Gubser, J. Maldacena, H. Ooguri,
Y. Oz, Phys. Rept. 323 (2000) 183; N. Drukker, D. J. Gross, 
Phys. Rev. D60 (1999) 125006; 
O. Aharony, hep-th/0212193;
F. Bigazzi, A. L. Cotrone, M. Petrini, A. Zaffaroni, hep-th/0303191. 
\bibitem{aw} J. J. Atick, E. Witten, Nucl. Phys. B310 (1988) 291.
\bibitem{toy} H. Neuberger, Nucl. Phys. B179 (1980) 253. 
\bibitem{do} B. Durhuus, P. Olesen, Nucl. Phys. B184 (1981) 461.
\bibitem{dk} M. Douglas, V. Kazakov, Phys. Lett. B319 (1993) 219.
\bibitem{wsph}  V. Kazakov, Phys. Lett. B105 (1981) 453.
\bibitem{gm} D. J. Gross, A. Matytsin, Nucl. Phys. B429 (1994) 50.
\bibitem{ren} R. A. Brandt, F. Neri, Phys. Rev. D24 (1981) 879.
\bibitem{mehta} P. J. Forrester, N. C. Snaith, J. J. M. Verbaarschot,
cond-mat/0303207. 
\bibitem{inst} H. Neuberger, Phys. Lett. B94 (1980) 199. 
\bibitem{thooft} G. 't Hooft, Nucl. Phys. B117 (1976) 519.
\end{thebibliography}
\end{document}